\documentclass[journal,twoside]{IEEEtran}
\usepackage[margin=2.2cm, centering]{geometry}
\usepackage{multirow}
\usepackage{booktabs}
\usepackage{lipsum}
\usepackage{amsfonts}
\usepackage{placeins}
\usepackage{array}
\usepackage{amsmath,cases,amssymb}
\usepackage{amsmath}
\usepackage{graphicx}
\usepackage{cite}
\usepackage{subfigure}
\usepackage{epsfig}
\usepackage{url}
\usepackage{algorithm}
\usepackage{algorithmic}
\usepackage{epstopdf}
\usepackage{balance}
\usepackage{color}
\usepackage{algorithm}
\DeclareGraphicsExtensions{.eps,.pdf}

\newcommand{\ds}{\displaystyle}

\newcommand{\st}{{\mathrm{s.t.}}}

\begin{document}

\title{Optimal User Pairing for Achieving Rate Fairness in Downlink NOMA Networks}
\author{Van-Phuc Bui, Phu X. Nguyen, Hieu V. Nguyen, Van-Dinh Nguyen, and Oh-Soon Shin\\
School of Electronic Engineering $\&$ Department of ICMC  Convergence Technology, \\ Soongsil University, Korea (E-mail: osshin@ssu.ac.kr)
\vspace{-0.5cm}
 \thanks{This research was supported in part by Basic Science Research Program through the National Research Foundation of Korea (NRF) funded by the Ministry of Education (No. 2017R1D1A1B03030436) and by the Ministry of Science and ICT (No. NRF-2017R1A5A1015596) and in part by the BK21 plus Program through NRF grant funded by the Ministry of Education (No. 31Z20150313339). 
}

}
\maketitle
\begin{abstract}
In this paper, a downlink non-orthogonal multiple access (NOMA) network is studied. We investigate the problem of jointly optimizing  user pairing and beamforming design to maximize the minimum rate among all users. The considered problem belongs to a difficult class of mixed-integer nonconvex optimization programming. We first relax the binary constraints and adopt sequential convex approximation method to solve the relaxed problem, which is guaranteed to converge at least to a locally optimal solution. Numerical results show that the proposed method attains higher rate fairness among users, compared with traditional beamforming solutions, i.e., random pairing NOMA and  beamforming systems.
\end{abstract}
\begin{IEEEkeywords}
Beamforming, convex optimization, non-orthogonal multiple access (NOMA).
\end{IEEEkeywords}

\section{Introduction}\label{sec:intro}
Non-orthogonal multiple access (NOMA) technique is being considered as a promising candidate of the fifth generation networks (5G) \cite{NOMA1,LuiProIEEE17}. By allowing multiple users to share the same time-frequency resources, NOMA adopts the successive interference cancellation (SIC) technique at users with better channel conditions that significantly improves the system performance in terms of the spectral efficiency and edge throughput of users with poorer channel conditions \cite{Dinh:JSAC:Dec2017}.

By following the principle of NOMA, two users with more distinct channel conditions should be paired \cite{NOMA2}. Thus, the optimal user pairing is crucial to further improve the performance of NOMA systems. In particular, random user pairing schemes were proposed in \cite{Dinh:JSAC:Dec2017, NOMA2 , DingTWC16,ChenCOMLL18}, where a near user  is randomly paired with one farther from the base station (BS). Nonetheless, a general criteria for dynamic user
pairing was not investigated. Recently, the optimal user pairing was studied in \cite{ZhuWCL18}, where a minimum rate constraint for all  NOMA
users is additionally imposed. All of the aforementioned NOMA works focused on either single-antenna scenarios and/or random user pairing schemes. In the power domain NOMA, the BS should allocate a higher portion of transmission power  to users with poor channel conditions from the viewpoint
of fairness, which will generate strong interference to near users. 

Motivated by the above discussion and shortcoming of the previous works, in this paper, we propose a dynamic user pairing and beamforming optimization problem for downlink (DL) NOMA systems from the perspective of fairness among users, which is inspired from the fact that the users with poor channel conditions may get a very low (even zero) throughput. The  dynamic user pairing is done by introducing binary variables, which results in a difficult class of mixed-integer nonconvex optimization problem. The exhaustive search over all possible cases of user pairing provides the optimal performance of the considered problem but with prohibitive complexity. The main contributions of this paper are two-fold: $(i)$ We formulate a general optimization problem of dynamic user pairing and beamforming design, which has not
been reported in the literature; $(ii)$ For efficient and practical implementations, we relax the binary constraints and propose a low-complexity iterative algorithm based on the inner approximation (IA) framework for its solution \cite{IA}, which only solves a sequence of simple convex programs. 
Numerical results are provided to confirm that our proposed scheme
is efficient in terms of the rate fairness.

\textit{Notation:} Vectors are denoted by bold lower-case letters. $\mathbf{w}^{H}$, $\mathbf{w}^{T}$ and $\mathbf{w}^{*}$   are the Hermitian transpose, normal transpose and conjugate  of a vector $\mathbf{w}$, respectively.
$\mathbb{E}\{ \cdot \}$ denotes the expectation and $\Re\{a\}$ returns the real part of a complex number $a$.
$\mathbb{R}$ and $\mathbb{C}$ denote the set of all real and complex numbers, respectively. $\|\cdot\|$ denotes   a vector's Euclidean norm.

\section{System Model and Problem Formulation}\label{sec:sys_model}

\subsection{System Model}

\begin{figure}[!t]
\centering
\includegraphics[width=0.46\textwidth,trim={0.0cm 0.0cm -0cm -0.0cm}]{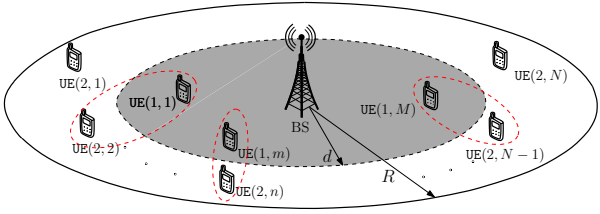}
\caption{An illustration of DL NOMA consisting of one BS and ($M+N$) users.}
\label{fig:SM:1}
\end{figure}

We consider  a downlink NOMA system consisting of one BS, a set of $M$ near users (UEs), denoted by $\mathcal{M} \triangleq \{1,2,\cdots, M\}$, and a set of $N$ far UEs, denoted by $\mathcal{N} \triangleq \{1,2,\cdots, N\}$, as illustrated in Fig. \ref{fig:SM:1}. In addition, $\mathtt{UE}(1,m)$ and $\mathtt{UE}(2,n)$ indicate the $m$-th UE with $m\in \mathcal{M}$ in inner-zone and the $n$-th UE with $n\in \mathcal{N}$ in outer-zone, respectively. For notational convenience, $\mathcal{G}\triangleq \{(1,1),\cdots,(1,M),(2,1),\cdots,(2,N)\}$ is defined as the set of all UEs. The BS is equipped with $L$ antennas while each user has a single antenna. Then, the received signal at  $\mathtt{UE}(i,j)$ can be expressed as
\begin{IEEEeqnarray}{rCl}
    y_{i, j} &=& \ds\sum_{(i',j')\in \mathcal{G}}\mathbf{h}_{i, j}^H\mathbf{w}_{i',j'}{x}_{i',j'} + {n}_{i,j}, \quad (i,j)\in\mathcal{G}
   \quad \label{eq:yij}
 \end{IEEEeqnarray}
where $\mathbf{h}_{i, j} \in\mathbb{C}^{L\times1}$ and $\mathbf{w}_{i,j} \in\mathbb{C}^{L\times 1}$ are the channel vector from the BS to $\mathtt{UE}(i,j)$ and the beamforming vector, respectively. $x_{i,j}$ with $\mathbb{E}\{|x_{i,j}|^2\}=1$  and ${n}_{i,j} \sim \mathcal{CN}(0,\sigma^2_{i,j})$ are the transmitted symbol and the additive white Gaussian noise (AWGN) at  $\mathtt{UE}(i,j)$, respectively.

In this paper, we assume that NOMA beamforming is exploited for pairs of two users, $\mathtt{UE}(1,m)$ and $\mathtt{UE}(2,n)$ $\forall m \in \mathcal{M}, \forall n \in \mathcal{N}$. To do so, we introduce new binary variables $\alpha _{m,n} \in \{0,1\}$  as
\begin{equation}
\alpha _{m,n} = \left\{
							\begin{array}{ll}
									1,\quad \text{if}\  \mathtt{UE}(1,m)\ \text{and}\ \mathtt{UE}(2,n)\ \text{are paired,}  \\
									0,\quad \text{otherwise}.
							\end{array}
			   \right.
\label{eq:CSIerrors}
\end{equation}
 In each pair, the SIC technique will be adopted at $\mathtt{UE}(1,m)$ by following NOMA principle. In particular, $\mathtt{UE}(1,m)$ first decodes the message intended to $\mathtt{UE}(2,n)$ and subtracts it before decoding $\mathtt{UE}(1,m)$'s message. On the other hand, $\mathtt{UE}(2,n)$ will decode its own message directly by treating $\mathtt{UE}(1,m)$'s message as noise. By defining $\boldsymbol{\alpha} \triangleq {[\alpha_{m,n}]}_{m\in \mathcal{M}, n\in \mathcal{N}}$, $\mathbf{w}_1 \triangleq {[\mathbf{w}_{1,m}]}_{m \in \mathcal{M}}$, $\mathbf{w}_2 \triangleq {[\mathbf{w}_{2,n}]}_{n\in \mathcal{N}}$ and $\mathbf{w} \triangleq {[\mathbf{w}_1^H \text{ } \mathbf{w}_2^H]^H}$,  the signal-to-interference-plus-noise ratio (SINR) at $\mathtt{UE}(1,m)$ and $\mathtt{UE}(2,n)$ can be written as
 \begin{IEEEeqnarray}{rCl}\label{eq:SINR}
{\gamma _{1,m}}(\mathbf{w},\boldsymbol{\alpha})   & =  &
 \frac{{{{\left| {\mathbf{h}_{{1,m}}^H{\mathbf{w}_{{1,m}}}} \right|}^2}}}{\Xi_{m} (\mathbf{w},\boldsymbol{\alpha})} ,\IEEEyessubnumber\label{eq:SINR1m}\\
\gamma _{2,n}(\mathbf{w},\boldsymbol{\alpha})&=&\underset{m\in\mathcal{M}}{\min} \Bigl\{ \frac{{\left| {\mathbf{h}_{{2,n}}^H{\mathbf{w}_{{2,n}}}} \right|}^2}{\Phi_{n} (\mathbf{w})}, \frac{{\left| {\mathbf{h}_{{1,m}}^H{\mathbf{w}_{{2,n}}}} \right|}^2}{\alpha_{m,n}\Psi_{m,n} (\mathbf{w})} \Bigr\}\quad \IEEEyessubnumber\label{eq:SINR2n}
 \end{IEEEeqnarray}
 where $\Xi_{m}(\mathbf{w},\boldsymbol{\alpha})$, $\Phi_{n} (\mathbf{w})$ and $\Psi_{m,n}(\mathbf{w})$ are defined as
    \begin{IEEEeqnarray}{lCl}
\Xi_{m}(\mathbf{w},\boldsymbol{\alpha})&=& \ds\sum _{m'\in \mathcal{M}\backslash \{m\}}\left|\mathbf{h}_{1,m}^H \mathbf{w}_{1,m'}\right|^2 \nonumber \\
&&+ \ds\sum _{n'\in \mathcal{N}}(1-\alpha_{m,n'})\left|\mathbf{h}_{1,m}^H \mathbf{w}_{2,n'}\right|^2 +\sigma_{1,m}^2, \nonumber \\
\Phi_{n} (\mathbf{w})&=& \ds\sum _{(i,j)\in \mathcal{G}\backslash \{(2,n)\}}\left|\mathbf{h}_{2,n}^H \mathbf{w}_{i,j}\right|^2 +\sigma_{2,n}^2 ,\nonumber\\
\Psi_{m,n}(\mathbf{w})&=& \ds\sum _{(i,j)\in \mathcal{G}\backslash\{(2,n)\}}\left|\mathbf{h}_{1,m}^H \mathbf{w}_{i,j}\right|^2 +\sigma_{1,m}^2.\nonumber
 \end{IEEEeqnarray}
In \eqref{eq:SINR2n},  $\frac{{\left| {\mathbf{h}_{{2,n}}^H{\mathbf{w}_{{2,n}}}} \right|}^2}{\Phi_{n} (\mathbf{w})}$ and $\frac{{\left| {\mathbf{h}_{{1,m}}^H{\mathbf{w}_{{2,n}}}} \right|}^2}{\alpha_{m,n}\Psi_{m,n} (\mathbf{w})}$ are the SINRs of $\mathtt{UE}(2,n)$ decoded at $\mathtt{UE}(2,n)$ and $\mathtt{UE}(1,m)$, respectively. Clearly, if $\alpha _{m,n}=0,\ \forall m,n$, then $\gamma _{2,n}(\mathbf{w},\boldsymbol{\alpha})= \frac{{\left| {\mathbf{h}_{{2,n}}^H{\mathbf{w}_{{2,n}}}} \right|}^2}{\Phi_{n} (\mathbf{w})}$.
\subsection{Problem Formulation}
From \eqref{eq:SINR}, the achievable throughput for $\mathtt{UE}(i,j)$ can be derived as
  \begin{IEEEeqnarray}{lCl} \label{Rate}
R_{i,j}(\mathbf{w},\boldsymbol{\alpha})=\log_2(1+\gamma_{i,j}(\mathbf{w},\boldsymbol{\alpha})), \quad (i,j)\in \mathcal{G}.
 \end{IEEEeqnarray}
Herein, we aim to maximize the minimum rate among all UEs, called max-min rate (MMR) for short. Accordingly, an optimization problem can be mathematically formulated as
 \begin{IEEEeqnarray}{lrCl} \label{maxminrate}
    &\max \limits_{\mathbf{w},\boldsymbol{\alpha}}&\quad & \min_{(i,j)\in \mathcal{G}} R_{i,j}(\mathbf{w},\boldsymbol{\alpha}) \IEEEyessubnumber\label{maxminratea}\\
&\st && \|\mathbf{w}\|^2 \leq P^{\max}_{\text{BS}},\quad \IEEEyessubnumber\label{maxminrateb}\\
&&&\alpha_{m,n} \in \{0,1\},\quad  \forall m \in \mathcal{M}, \forall n\in \mathcal{N},\quad \IEEEyessubnumber\label{maxminratec}\\
&&&\sum\limits_{n\in\mathcal{N}}\alpha_{m,n} \leq 1,\quad \forall m \in\mathcal{M}\IEEEyessubnumber\label{maxminrated},\\
&&&\sum\limits_{m\in\mathcal{M}}\alpha_{m,n} \leq 1,\quad \forall n\in \mathcal{N}\IEEEyessubnumber\label{maxminratee}
\end{IEEEeqnarray}
 where \eqref{maxminrateb} represents the transmit power constraint at the BS, $P_{\text{BS}}^{\text{max}}$ and constraints \eqref{maxminratec}, \eqref{maxminrated}, \eqref{maxminratee} establish the criteria for user pairing. Specifically, constraints \eqref{maxminrated} and \eqref{maxminratee} ensure that each $\mathtt{UE}(i,j)$ can opportunistically pair to  one UE only. We can see that \eqref{maxminratea} is nonsmooth and nonconcave and \eqref{maxminratec} corresponds to binary constraints, leading to a mixed-integer nonconvex optimization  of problem \eqref{maxminrate}. Thus, problem \eqref{maxminrate} is intractable and it may not be possible to convert the problem into an equivalent convex one. Following some recent studies on wireless communication system designs \cite{Nguyen:TCOM:17,NguyenJSAC18}, we aim at solving \eqref{maxminrate} based on the application of IA method, which efficiently provides a locally optimal solution with low complexity.

\section{Proposed Iterative Algorithm }\label{section_alg}

The main difficulty of solving \eqref{maxminrate} is to handle the binary constraints \eqref{maxminratec}. To overcome this issue, we first relax $\alpha_{m,n}\in\{0,1\}$ to $0\leq\alpha_{m,n}\leq 1$ and rewrite \eqref{maxminratec} as
 \begin{subequations}\label{maxminrate2}
   \begin{IEEEeqnarray}{lrCl} 
  & \max \limits_{\mathbf{w},\boldsymbol{\alpha}}&\quad & \min \limits_{(i,j)\in \mathcal{G}} \gamma_{i,j}(\mathbf{w},\boldsymbol{\alpha}) \label{maxminrate2a}\\
&\st &&0\leq\alpha_{m,n}\leq 1,\quad \forall m \in \mathcal{M}, \forall n\in \mathcal{N},\quad \label{maxminrate2b}\\
&&&\eqref{maxminrateb}, \eqref{maxminrated}, \eqref{maxminratee}\quad \label{maxminrate2c}
\end{IEEEeqnarray}
 \end{subequations}
where the optimal solutions of MMR problem and max-min SINR problem are identical. We can observe that the feasible sets are convex (quadratic and linear constraints), and thus only the objective function \eqref{maxminrate2a} remains nonconcave.
In order to apply the IA method, we introduce a new variable $\beta$ to re-express \eqref{maxminrate2} equivalently as
\begin{subequations}\label{maxminrate3}
   \begin{IEEEeqnarray}{lrCl} 
  & \max \limits_{\mathbf{w},\boldsymbol{\alpha},{\beta}}&\quad &  1/\beta \label{maxminrate3a}\\
&\st &&1/\beta \leq \gamma_{i,j},\quad (i,j)\in \mathcal{G},\label{maxminrate3b}\\
&&& \eqref{maxminrateb}, \eqref{maxminrated}, \eqref{maxminratee}, \eqref{maxminrate2b}.\quad \label{maxminrate3c}
\end{IEEEeqnarray}
 \end{subequations}
We remark that the equivalence between \eqref{maxminrate2} and \eqref{maxminrate3} is guaranteed due to the fact that constraints \eqref{maxminrate3b} must hold with equality at optimum \cite{NguyenJSAC18}.

\underline{\textit{Concavity of \eqref{maxminrate3a}}:} The function $1/\beta$ is convex and thus its first order approximation at a feasible point $\beta^{(\kappa)}$ found at iteration $\kappa$ is given by
\[\frac{1}{\beta} \geq \frac{2}{\beta^{(\kappa)}} - \frac{\beta}{(\beta^{(\kappa)})^2}\]
which is a linear function.

\underline{\textit{Convexity of \eqref{maxminrate3b}}:} Here, we consider two  cases of $i\in\{1,2\}$ due to  different structures of the SINR functions.\vspace{5pt}

\noindent \underline{($i$) With $i=1$}: constraint \eqref{maxminrate3b} is $1/\beta \leq \gamma_{1,j}$, which can be revised as
\begin{IEEEeqnarray}{cCl}\label{ctSRNRzone1}
\Xi_m(\mathbf{w},\boldsymbol{\alpha}) \leq \left| \mathbf{h}_{1,m}^H \mathbf{w}_{1,m}\right|^2 \beta.
\end{IEEEeqnarray}
Next, we introduce new variables $\boldsymbol{\tau} \triangleq\{\tau_{m,n}\}_{m\in \mathcal{M}, n \in \mathcal{N}}$ to decompose \eqref{ctSRNRzone1} into the following two constraints:
\begin{subnumcases}{\label{problemLBequi:dequi} \eqref{ctSRNRzone1}\Leftrightarrow}
	  \left|\mathbf{h}_{1,m}^H\mathbf{w}_{2,n'}\right|^2\leq \tau_{m,n'}\ , \ m\in \mathcal{M}, n' \in \mathcal{N}, & \IEEEyessubnumber\label{tau}\\
   \Xi_m(\mathbf{w},\boldsymbol{\alpha}, \boldsymbol{\tau}) \leq \left| \mathbf{h}_{1,m}^H \mathbf{w}_{1,m}\right|^2 \beta\  &\IEEEyessubnumber\label{ctSRNRzone1:d4}
\end{subnumcases}
where $\Xi_{m}(\mathbf{w},\boldsymbol{\alpha},\boldsymbol{\tau})= \ds\sum _{m'\in \mathcal{M}\backslash \{m\}}\left|\mathbf{h}_{1,m}^H \mathbf{w}_{1,m'}\right|^2 
+ \ds\sum _{n'\in \mathcal{N}}(1-\alpha_{m,n'})\tau_{m,n'} +\sigma_{1,m}^2.$ Since \eqref{tau} is convex constraint, we only need to handle the non-convexity of \eqref{ctSRNRzone1:d4}. Constraint \eqref{ctSRNRzone1:d4} is innerly convexified as
\begin{IEEEeqnarray}{cCl}\label{eq:Xi}
\widehat\Xi_m^{(\kappa)}(\mathbf{w},\boldsymbol{\alpha},\boldsymbol{\tau}) \leq f_{1,m}^{(\kappa)}(\mathbf{w})\beta  , \quad m\in\mathcal{M}.
\end{IEEEeqnarray}
where $\widehat\Xi_m^{(\kappa)}(\mathbf{w},\boldsymbol{\alpha},\boldsymbol{\tau})$ is a convex upper bound of $\Xi_m(\mathbf{w},\boldsymbol{\alpha}, \boldsymbol{\tau})$ by using \cite[Eq. (B.1)]{NguyenJSAC18}:
\begin{IEEEeqnarray}{lCl}
\widehat\Xi_m^{(\kappa)}(\mathbf{w},\boldsymbol{\alpha},\boldsymbol{\tau}) \triangleq \sum\limits_{m'\in \mathcal{M}\backslash \{m\}}\left|\mathbf{h}_{1,m}^H \mathbf{w}_{1,m'}\right|^2  + \sigma_{1,m}^2 +\nonumber \\
\ds\sum _{n'\in \mathcal{N}}\Bigl(\frac{1}{2}\frac{1-\alpha_{m,n'}^{(\kappa)}}{\tau_{m,n'}^{(\kappa)}}\tau_{m,n'}^2 + \frac{1}{2}\frac{\tau_{m,n'}^{(\kappa)}}{1-\alpha_{m,n'}^{(\kappa)}}(1-\alpha_{m,n'})^2 \Bigr) \nonumber
\end{IEEEeqnarray}
and $f_{1,m}^{(\kappa)}(\mathbf{w})$ is a lower bound of the convex function $\left| \mathbf{h}_{1,m}^H \mathbf{w}_{1,m}\right|^2$ given as \cite[Eq. (22)]{NguyenJSAC18}:
\begin{IEEEeqnarray}{lll}\label{f1}
{\left| \mathbf{h}_{1,m}^H \mathbf{w}_{1,m}\right|^2} \geq&  {2\Re\{(\mathbf{h}_{1,m}^H \mathbf{w}_{1,m}^{(\kappa)})^H(\mathbf{h}_{1,m}^H \mathbf{w}_{1,m})\}} \nonumber\\
& - {|\mathbf{h}_{1,m}^H \mathbf{w}_{1,m}^{(\kappa)}|^2} \quad \triangleq f_{1,m}^{(\kappa)}(\mathbf{w}).\nonumber
\end{IEEEeqnarray}

\noindent \underline{($ii$) With $i=2$}: constraint \eqref{maxminrate3b} is replaced by
\begin{subnumcases}{\label{betazone2} }
	   1/\beta\leq \frac{{| {\mathbf{h}_{{2,n}}^H{\mathbf{w}_{{2,n}}}} |}^2}{\Phi_{n} (\mathbf{w})}, \quad \forall n\in\mathcal{N}, &\qquad \IEEEyessubnumber\label{betazone2a}\\
    1/\beta  \leq \frac{{| {\mathbf{h}_{{1,m}}^H{\mathbf{w}_{{2,n}}}} |}^2}{\alpha_{m,n}\Psi_{m,n} (\mathbf{w})}, \quad \forall m\in\mathcal{M}, \forall n\in\mathcal{N}.\IEEEyessubnumber\label{betazone2b} &\qquad 
\end{subnumcases}
Similarly to \eqref{ctSRNRzone1:d4}, constraints in \eqref{betazone2} are innerly covexified as
\begin{subnumcases}{\label{betazone2Convex} }
	   \Phi_{n} (\mathbf{w})\leq f_{2,n}^{(\kappa)}(\mathbf{w})\beta,\quad \forall n\in\mathcal{N}, & \IEEEyessubnumber\label{betazone2aConvex}\\
   \Psi_{m,n}(\mathbf{w}) \leq  \tilde{f}_{m,n}^{(\kappa)}(\mathbf{w},\boldsymbol{\alpha})\beta, \quad \forall m\in\mathcal{M}, \forall n\in\mathcal{N}&\IEEEyessubnumber\label{betazone2bConvex} \qquad 
\end{subnumcases}
where $f_{2,n}^{(\kappa)}(\mathbf{w})$ is a lower bound of $| {\mathbf{h}_{{2,n}}^H{\mathbf{w}_{{2,n}}}}|^2$ given as:
\begin{IEEEeqnarray}{lll}\label{f2}
 f_{2,n}^{(\kappa)}(\mathbf{w}) \triangleq {2\Re\{(\mathbf{h}_{2,n}^H \mathbf{w}_{2,n}^{(\kappa)})^H(\mathbf{h}_{2,n}^H \mathbf{w}_{2,n})\}} - {|\mathbf{h}_{2,n}^H \mathbf{w}_{2,n}^{(\kappa)}|^2} \nonumber
\end{IEEEeqnarray}
and $\tilde{f}_{m,n}^{(\kappa)}(\mathbf{w},\boldsymbol{\alpha})$ is a lower bound of ${\left| {\mathbf{h}_{{1,m}}^H{\mathbf{w}_{{2,n}}}} \right|}^2/\alpha_{m,n}$ derived by using \cite[Eq. (38)]{NguyenJSAC18}:
\begin{IEEEeqnarray}{lll}
\tilde{f}_{m,n}^{(\kappa)}(\mathbf{w},\boldsymbol{\alpha}) \triangleq &&\frac{2\Re\{(\mathbf{h}_{1,m}^H \mathbf{w}_{2,n}^{(\kappa)})^H(\mathbf{h}_{1,m}^H \mathbf{w}_{2,n})\}}{\alpha_{m,n}^{(\kappa)}}\nonumber\\
&&-\frac{|\mathbf{h}_{1,m}^H\mathbf{w}_{2,n}^{(\kappa)}|^2}{{(\alpha_{m,n}^{(\kappa)})}^2}(\alpha_{m,n}) .\nonumber\end{IEEEeqnarray}
We should note that constraints \eqref{eq:Xi} and \eqref{betazone2Convex} can be expressed as second-order cone (SOC) constraints, which are obviously convex ones.

Summing up, problem \eqref{maxminrate2} can be approximated as the following  convex program at iteration $(\kappa +1)$:
   \begin{IEEEeqnarray}{lrCl} \label{MaxMinFinal}
  & \max \limits_{\mathbf{w},\boldsymbol{\alpha},{\beta},\boldsymbol{\tau}}&\quad &  \eta \triangleq \frac{2}{\beta^{(\kappa)}} - \frac{\beta}{(\beta^{(\kappa)})^2} \IEEEyessubnumber\label{MaxMinFinala}\\
&\st && \eqref{maxminrateb}, \eqref{maxminrated}, \eqref{maxminratee},\eqref{maxminrate2b},
\eqref{tau}, \eqref{eq:Xi}, \eqref{betazone2Convex}.\quad \IEEEyessubnumber\label{MaxMinFinalb}
\end{IEEEeqnarray}
We have numerically observed that some values of $\alpha_{m,n}$ are very close to binary but not exactly binary values at the optimum. This makes \eqref{maxminrate} infeasible. Therefore, we further introduce the rounding function after obtaining the optimal solution of problem \eqref{MaxMinFinal} as
\begin{IEEEeqnarray}{lCl} \label{rounding}
\alpha_{m,n}^\star = \Big\lfloor \alpha_{m,n}^{(\kappa)} +\frac{1}{2} \Big\rfloor, \quad m\in\mathcal{M},n\in\mathcal{N}.
\end{IEEEeqnarray}
The proposed algorithm is summarized in Algorithm \ref{alg_1}. Specifically, Algorithm \ref{alg_1} consists of two phases: In phase 1, we successively solve \eqref{MaxMinFinal} to achieve $(\mathbf{w}^{(\star)}, \boldsymbol{\alpha}^{(\star)}, \beta^{(\star)}, \boldsymbol{\tau}^{(\star)})$. In phase 2,  we first use the rounding function (\ref{rounding}) to force $\boldsymbol{\alpha}$ into the nearest Boolean values, and then resolve problem \eqref{MaxMinFinal} for a fixed value of $\boldsymbol{\alpha}$ to find the optimal solution $\mathbf{w}^{(\star)}$. Moreover, due to the fact that the IA method
is employed, Algorithm \ref{alg_1} converges to a stationary point, which also satisfies  the Karush-Kuhn-Tucker (KKT) conditions
of \eqref{maxminrate2}. The detailed proof can be done by following the same steps in \cite{IA,NguyenJSAC18}.

\begin{algorithm}[t]
\begin{algorithmic}[1]
\protect\caption{Proposed Iterative Algorithm to Solve  \eqref{maxminrate2}}
\label{alg_1}
\global\long\def\algorithmicrequire{\textbf{Initialization:}}
\REQUIRE  Set $\kappa=0$ and generate feasible initial points $(\mathbf{w}^{(0)}, \boldsymbol{\alpha}^{(0)},{\beta}^{(0)}, \boldsymbol{\tau}^{(0)})$.\\
\global\long\def\algorithmicrequire{\textbf{Phase 1:}}
\REQUIRE
\REPEAT
\STATE Solve the convex program \eqref{MaxMinFinal} to compute the optimal solution $(\mathbf{w}^\star,\boldsymbol{\alpha}^\star,{\beta}^\star, \boldsymbol{\tau}^\star)$.
\STATE Update\ \ $  (\mathbf{w}^{(\kappa+1)},\boldsymbol{\alpha}^{(\kappa+1)},{\beta}^{(\kappa+1)}, \boldsymbol{\tau}^{(\kappa+1)}) = (\mathbf{w}^\star,\boldsymbol{\alpha}^\star,{\beta}^\star,\boldsymbol{\tau}^\star)$.
\STATE Set $\kappa=\kappa+1.$
\UNTIL Convergence\\
\STATE \textbf{Output-1:}  $(\mathbf{w}^\star,{\beta}^\star, \boldsymbol{\tau}^\star) =  (\mathbf{w}^{(\kappa)},{\beta}^{(\kappa)}, \boldsymbol{\tau}^{(\kappa)})$,\\
\quad \quad \quad \quad \quad $\boldsymbol{\alpha}^\star$ is updated by  (\ref{rounding}).\\
\global\long\def\algorithmicrequire{\textbf{Phase 2:}}
\REQUIRE
\STATE Run steps 1- 5 again to find beamformers $\mathbf{w}$ with fixed $\boldsymbol{\alpha}$.\\
\STATE \textbf{Output-2:} ($\mathbf{w}^\star, \boldsymbol{\alpha}^\star$).
\end{algorithmic} \end{algorithm}

\textit{Complexity analysis}: Since problem \eqref{MaxMinFinal} has $ x = (2(M+N) + 3MN + 1)$ quadratic and linear constraints and $ y = L(M+N)+2MN+1$ optimization variables , the per-iteration complexity  of solving \eqref{MaxMinFinal} is $\mathcal{O}\big(x^{2.5}(y^2 + x)\big)$ \cite{sedumi}.

\section{Numerical Results} \label{sec:simulation}
We consider a small-cell network serving 8 UEs with $M$ = 3 UEs in inner-zone  and $N$ = 5 UEs in outer-zone. Other important parameters are included in Table I. Algorithm 1 is terminated when the increase in the objective value between two consecutive iterations is less than $10^{-3}$.
\begin{table}[t]
\caption{SIMULATION PARAMETERS}
	\label{parameter}
	\centering
	{
		\begin{tabular}{l|l}
		\hline\hline
				Parameters & Value \\
		\hline
		    	Bandwidth                             &  20 [MHz] \\
				Noise power density & -174 [dBm/Hz] \\
				Path loss from the  BS to a  UE, $\sigma_{\mathsf{PL}}$   & 140 + 37.6$\log_{10}(d)$ [dB]\\
				Shadowing standard deviation & 8 [dB] \\
				Radius of the cell $(R)$ &  100 [m]\\
				Coverage of near UEs $(d)$  & 50 [m]\\
				Distance between BS and the nearest UE & $\geq$ 5 [m]\\
		\hline		   				
		\end{tabular}}
\end{table}	

\begin{figure}[t]
		\includegraphics[width=0.45\textwidth]{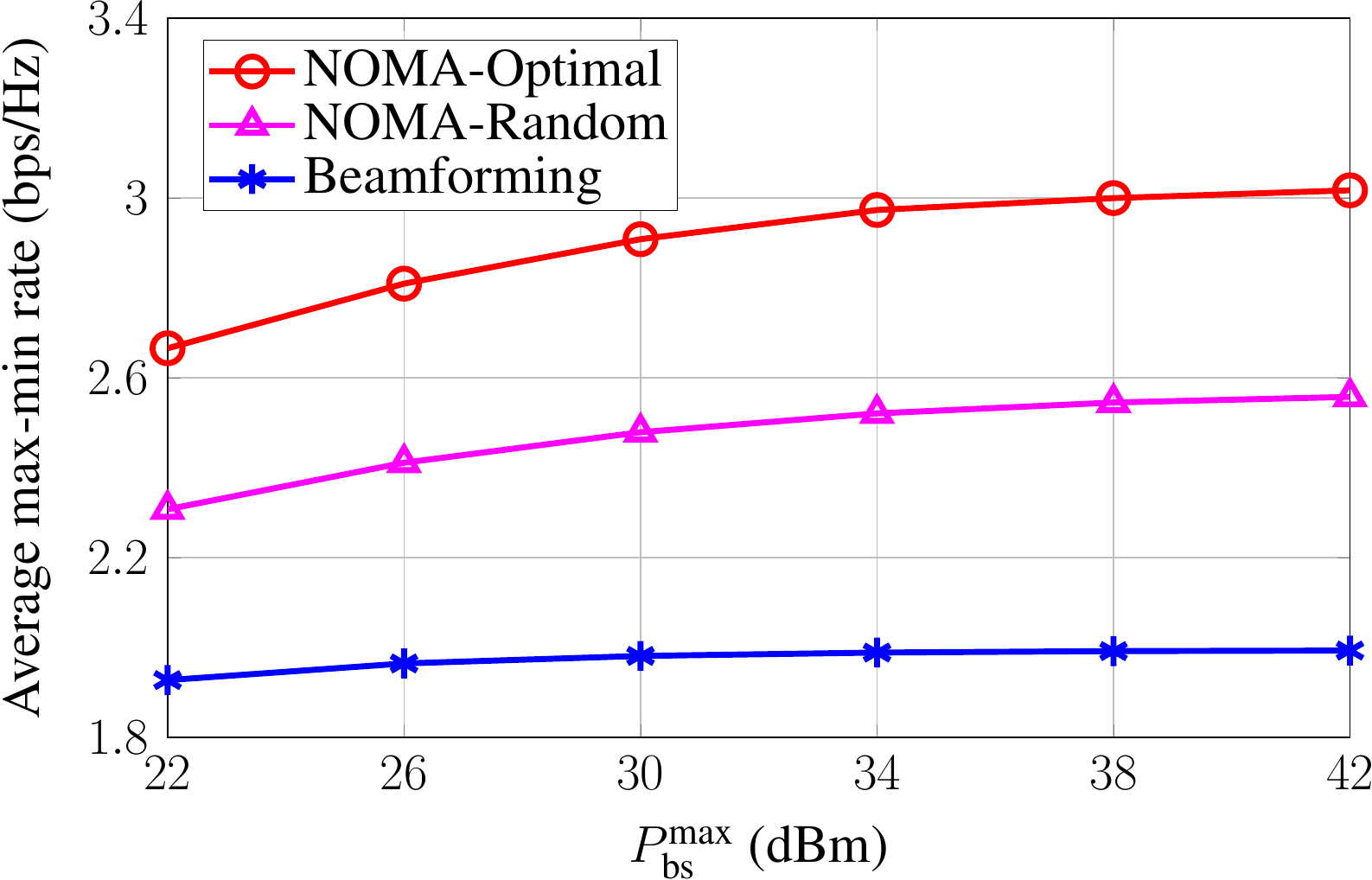}
		\caption{Average max-min rate performance versus $P_{\mathrm{BS}}^{\max}$ with $L = 6$.}\label{fig:CDFMIMO}
		\label{fig:compare}
	\end{figure}

Fig.~\ref{fig:compare} depicts the averaged MMR performance of our proposed algorithm (NOMA-Optimal) with two other resource allocation schemes as a function of $P_{\mathrm{BS}}^{\max}$. The solution of NOMA-Random \cite{Dinh:JSAC:Dec2017} is found by using Algorithm 1 with $\alpha_{m,n}$ being randomly chosen. The results are averaged over 100 random channel realizations. As expected, the MMR of Algorithm 1 is  higher  than that of NOMA-Random and beamforming schemes, of which the gaps are about 0.5 bps/Hz and 1 bps/Hz, respectively. This further demonstrates the effectiveness of jointly optimizing user pairing and beaforming design.

\begin{figure}[t]
		\includegraphics[width=0.45\textwidth]{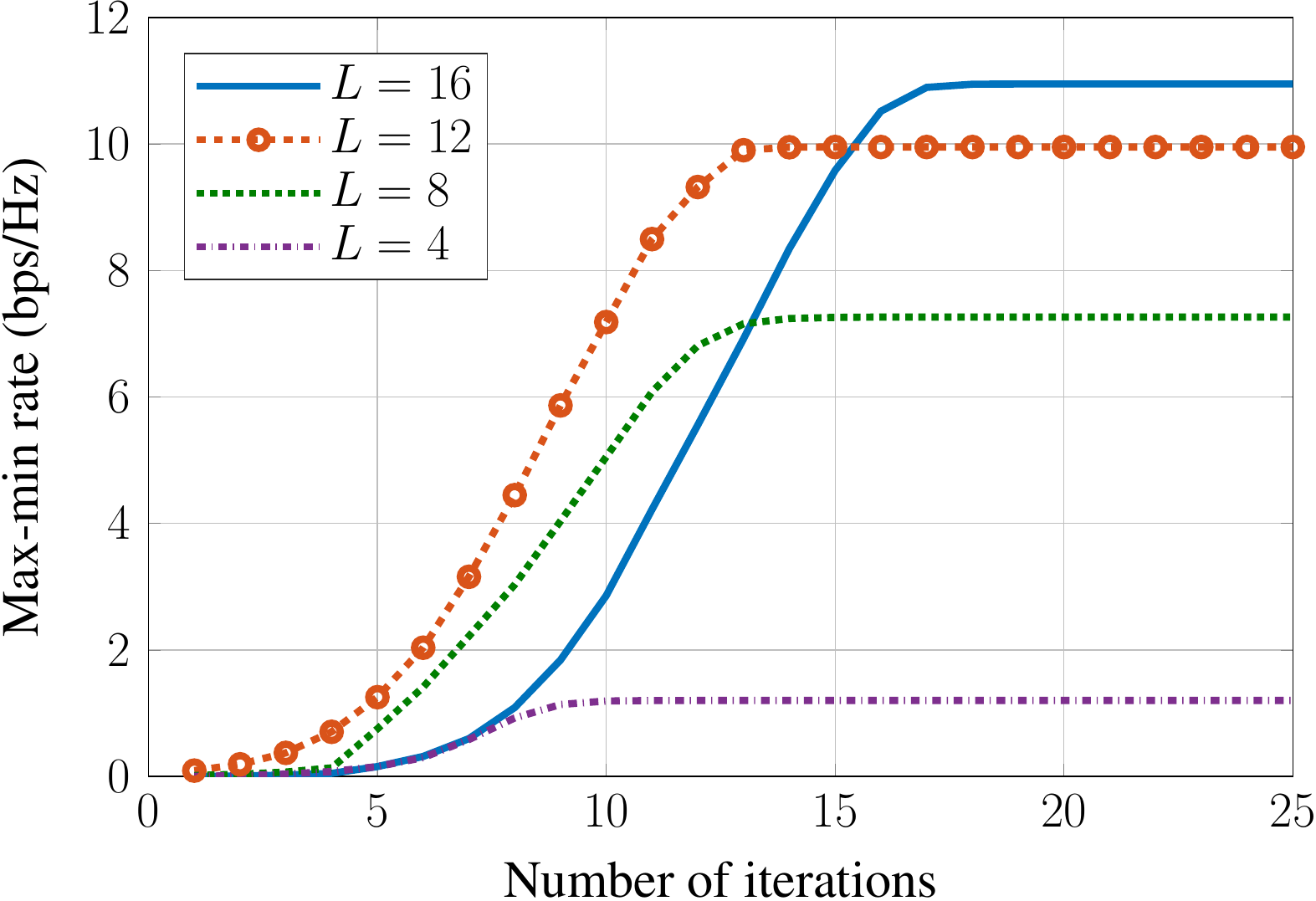}
		\caption{Convergence of Alg. 1 for one channel realization with $P_{\text{BS}}^{\text{max}} = 30\text{dBm}$. }
		\label{fig:convergence}
	\end{figure}

In Fig.~\ref{fig:convergence}, we illustrate the convergence behavior of Algorithm 1  for different number of antennas at the BS. We can see that increasing the number of antennas $L$ requires more number of iterations to converge. Nevertheless, it requires a few iterations to achieve the optimal solution, i.e., about 17 iterations for $L = 16$.


\section{Conclusions}\label{sec:conclusion}
This paper  studied a  DL NOMA network, where the optimal user pairing is investigated. We formulated a max-min rate optimization problem to jointly optimize  user pairing and beamforming design, and then designed an efficient iterative algorithm based on the IA method to solve it.  The effectiveness of the proposed algorithm has been demonstrated by the numerical results.

\bibliographystyle{IEEEtran}
\bibliography{Journal}
\end{document}